\documentclass[prl,10pt,floatfix,amsmath,twocolumn,superscriptaddress]{revtex4-1}
\usepackage[ansinew]{inputenc}
\usepackage{graphicx}
\usepackage{pst-all}
\usepackage{color}
\usepackage{dcolumn}
\usepackage{amsmath}
\usepackage{amsthm}
\usepackage{bm}
\usepackage{layout}
\usepackage{float}
\usepackage{txfonts}
\usepackage{amsfonts}
\usepackage{amssymb}%
\usepackage[ansinew]{inputenc}
\usepackage{graphicx}
\usepackage{amsmath}
\usepackage{amsthm}
\usepackage{bm}
\usepackage{layout}
\usepackage{float}
\usepackage{amsfonts}
\usepackage{amssymb}
\usepackage{array}

\usepackage{bm}
\usepackage{graphicx}
\usepackage[colorlinks=true,linkcolor=blue,
pagecolor=blue,filecolor=blue,
menucolor=blue,urlcolor=blue,
citecolor=blue,anchorcolor=blue]{hyperref}%

\begin{document}

\title{Phononic Weyl Nodal Straight Lines in High-Temperature Superconductor MgB$_2$}

\author{Qing Xie$^{\textcolor{blue}{\ddagger}}$}
\affiliation{Shenyang National Laboratory for Materials Science,
Institute of Metal Research, Chinese Academy of Sciences, Shenyang
110016, China} \affiliation{University of Chinese Academy of
Sciences, Beijing 100049, China}

\author{Jiangxu Li$^{\textcolor{blue}{\ddagger}}$}
\affiliation{Shenyang National Laboratory for Materials Science,
Institute of Metal Research, Chinese Academy of Sciences, Shenyang
110016, China} \affiliation{School of Materials Science and
Engineering, University of Science and Technology of China, 110016
Shenyang, Liaoning, China}

\author{Min Liu}
\affiliation{Shenyang National Laboratory for Materials Science,
Institute of Metal Research, Chinese Academy of Sciences, Shenyang
110016, China} \affiliation{School of Materials Science and
Engineering, University of Science and Technology of China, 110016
Shenyang, Liaoning, China}

\author{Lei Wang}
\affiliation{Shenyang National Laboratory for Materials Science,
Institute of Metal Research, Chinese Academy of Sciences, Shenyang
110016, China} \affiliation{School of Materials Science and
Engineering, University of Science and Technology of China, 110016
Shenyang, Liaoning, China}

\author{Dianzhong Li}
\affiliation{Shenyang National Laboratory for Materials Science,
Institute of Metal Research, Chinese Academy of Sciences, Shenyang
110016, China}

\author{Yiyi Li}
\affiliation{Shenyang National Laboratory for Materials Science,
Institute of Metal Research, Chinese Academy of Sciences, Shenyang
110016, China}

\author{ Xing-Qiu Chen}
\email{xingqiu.chen@imr.ac.cn} \affiliation{Shenyang National
Laboratory for Materials Science, Institute of Metal Research,
Chinese Academy of Sciences, Shenyang 110016, China}

\date{\today}

\begin{abstract}

Based on first-principles calculations, we predict that the
superconducting MgB$_2$ with a AlB$_2$-type centrosymmetric lattice
host the so-called phononic topological Weyl nodal lines (PTWNLs) on
its bulk phonon spectrum. These PTWNLs can be viewed as countless
Weyl points (WPs) closely aligned along the straight lines in the
$-$H-K-H direction within the three-dimensional Brillouin zone (BZ).
Their topological non-trivial natures are confirmed by the
calculated Berry curvature distributions on the planes perpendicular
to these lines. These lines are highly unique, because they exactly
locate at the high-symmetry boundary of the BZ protected by the
mirror symmetry and, simultaneously, straightly transverse the whole
BZ, in different from known classifications including nodal rings,
nodal chains or nets, and nodal loops. On the (10$\bar{1}$0) crystal
surface, the PTWNLs-induced drumhead-like non-trivial surface states
appear within the rectangular area confined by the projected lines
of the PTWNLs with opposite chirality. Moreover, when the mirror
symmetry is broken, the double-degenerate PTWNLs are further lifted
to form a pair WPs with opposite chirality. Our results pave the
ways for future experimental study on topological phonons on MgB$_2$
and highlights similar results in a series of isostructural
AlB$_2$-type metallic diborides.

\end{abstract}

%\pacs{63.20.Dj, 03.65.Vf}

\maketitle

%{\bf\color{blue} ================================== }

%
%   introduction
%
%\emph{Introduction}
As accompanying with extensive studies of topological insulators,
topological semimetals and even topological superconductors in their
electronic structures\cite{Kane2015b,Hasan2010,Qi2011}, topological
phonons \cite{He2016,Mousavi2015,Liuduan2016,Zhanglifa2015,Lu.L2013}
have been most recently attracted attentions in condensed matter
physics and material sciences. This is mainly because the
topological phononic states will be interesting for potential
applications. In different from various electronic fermions,
phonons, as one of bosons, are not limited by the Pauli exclusion
principle. This fact demonstrates that the whole frequency zone of
phonon spectrum can be physically probed. In similarity to
topological properties of electrons, the topological effects of
phonons can induce the one-way edge phonon states or topologically
protected surface phononic states. Physically, these states will
conduct phonon with little or no
scattering\cite{He2016,Mousavi2015}, suggesting possible
applications for designing phononic circuits\cite{Liuduan2016}. For
instance, utilizing the one-way edge phonon states an ideal phonon
diode \cite{Liuduan2016} was proposed with a fully 100\% efficiency
in a multi-terminal transport system. It was even theoretically
uncovered that the chirality of topological phonons excited by
polarized photons can be detected by a valley phonon Hall effect in
monolayer hexagonal lattices \cite{Zhanglifa2015}. In particular, it
needs to be emphasized that in similarity to various fermions of
electrons, the exciting progresses of the bosons (vibrational
phonons) have been also predicted \cite{Lu.L2013} or observed in the
3D momentum space of solid crystals with the topological vibrational
states, such as Dirac, Weyl and line-node phonons in photonic
crystals with macroscopic acoustic systems of kHz frequency
\cite{Lu.L2013,Lu.L2015,Huber.S2016,Prodan.E2009,Chen.B2014,Yang.Z2015,
Wang.P2015,Xiao.M2015,Nash.L2015,Susstrunk2015,
Mousavi2015,Fleury2016,Rocklin2016,He2016,Susstrunk2016,Lifeng2017}
and theoretically predicted doubly-Weyl phonons in transition-metal
monosilicides with atomic vibrations at THz frequency
\cite{Zhang.T2017}. Most recently, also in a series of WC-type
family of materials (i.e., ZrS, ZrSe and ZrTe) with atomic vibration
at THz frequency, the single topological Weyl phonons are predicted
and they exist opening topological arcs of surface phonons,
connecting pairs of surface-projected Weyl points with opposite
chirality\cite{Chen2017}. However, to date no phononic topological
Weyl nodal lines have been reported in macroscopic acoustic systems
at kHz frequency or in atomic vibrational periodic lattices at THz
frequency.

\begin{figure}
\includegraphics[width=0.47\textwidth]{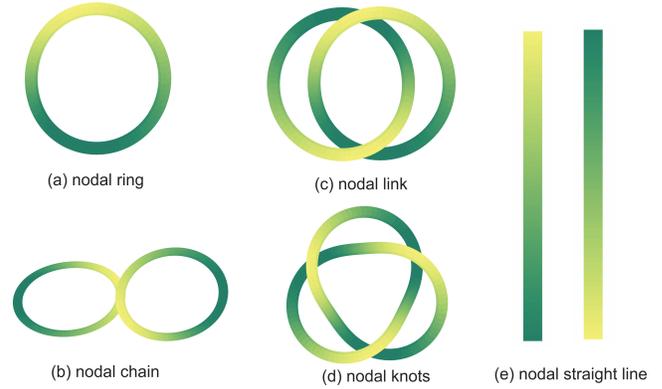}
\caption{Known main classifications of topological nodal lines: (a)
isolated nodal ring, (b) nodal chain, (c) nodal link, (d) nodal
knot, and (e) nodal straight lines extending the whole BZ in an
one-way direction. } \label{fig:nl}
\end{figure}

\begin{figure*}\begin{center}
\includegraphics[width=0.9\textwidth]{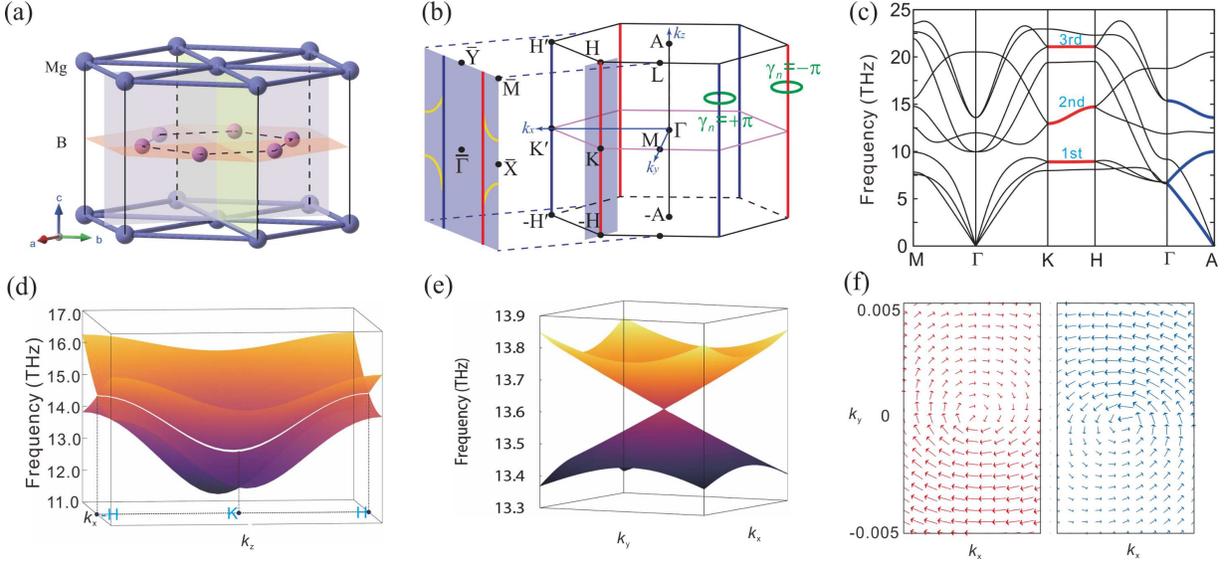}
\caption{(a) Crystal structure of MgB$_2$. Three shadowed planes are
the mirror planes. (b) The Brillouin zone and (10$\bar{1}$0)
surface. The PTWNLs transverse the whole BZ along both the $-H$ to
$H$ and the $-H^\prime$ to $H^\prime$ directions. Red (blue) lines
have negative (positive) chirality. Each PTWNL extends through the
whole BZ 1 times along one of the direction and carries a
topological charge \emph{Q} = $\gamma_n$/$\pi$ = $\pm$1 using the
input of the Berry phase. The surface states at about 14.2 THz are
schematically depicted on the surface. (c) The phonon dispersion of
MgB$_2$ along the high-symmetry momentums paths in the BZ. Three red
thick bands (as marked by the 1st, 2nd, and 3rd) along the H-K
direction correspond to double-degenerate PTWNLs, whereas three blue
thick double-generate bands along the $\Gamma$-A direction are
topologically trivial. (d) Three-dimensional phononic bands around
the 2nd PTWNL on the shaded \emph{k}-plane across the $-$H-K-H
direction as shown in the 3D BZ in the panel (b). (e)
Three-dimensional phononic bands at any WP on any PTWNL. Here, we
visualize it by selecting a node on the 2nd PTWNL at $k_{z}$ = 0.2.
(f) Berry curvature distributions on the ${xy}$ plane around two
nodal points of the 2nd PTWNLs at (1/3,1/3,0.2) and (-1/3,2/3,0.2),
respectively.} \label{fig:structure}
\end{center}
\end{figure*}

\begin{figure*}[hbt]
\centering \vspace{0.1cm}
\includegraphics[width=0.95\textwidth]{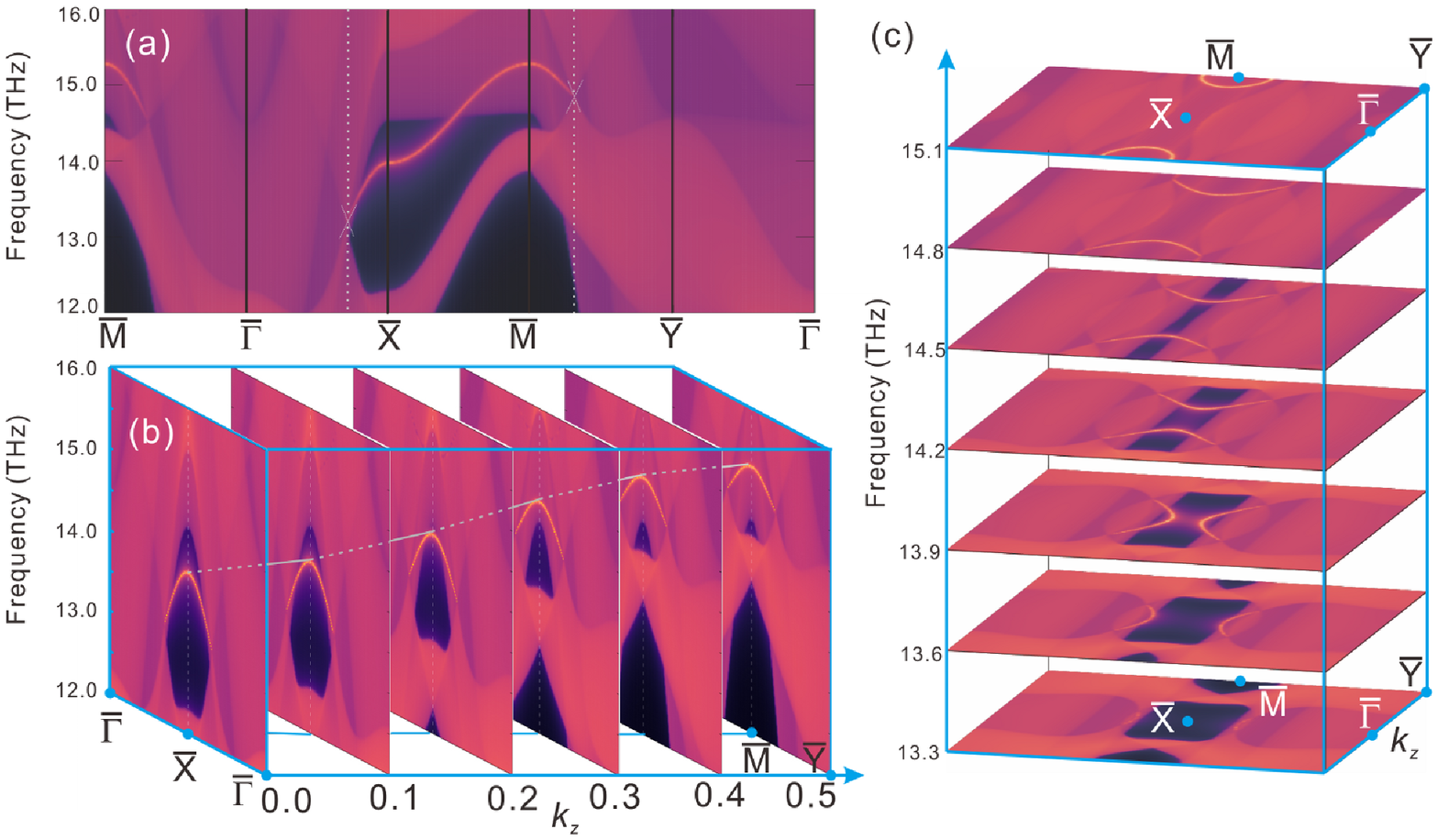}
\caption{Phononic states on the (10$\bar{1}$0) surface of MgB$_2$.
(a) The surface phonon spectrum along high symmetry lines. (b) The
evolution of the surface phonon spectra and the topologically
protected phononic non-trivial surface states by increasing the
$k_z$ value along the $\bar{\Gamma}$ - $\bar{X}$ - $\bar{\Gamma}$
direction. (c) The frequency-dependent evolution of the phononic
surface states of the (10$\bar{1}$0) BZ as defined in Fig.
\ref{fig:structure}(b) to show the change of the phononic drumhead
non-trivial surface states in the rectangular region confined by the
projected lines of two PTWNLs with opposite chirality.}
\label{fig:ss}
\end{figure*}

It needs to be emphasized that topological nodal lines
\cite{Fang2016} have been already and extensively studied in
electronic band structures, optical lattices, photonic and acoustic
systems. It has been found that symmetries, i.e., both $\mathcal{P}$
and $\mathcal{T}$ ones, can protect the presence of various nodal
lines. Once these symmetries are broken, nodal lines would be gapped
out or form Dirac points (DPs) or WPs. To date, nodal lines have
been classified into various categories, such as (\emph{i}) isolated
closed nodal rings (Fig.~\ref{fig:nl}(a)) in alkaline earth metals
of hcp Be and Mg ~\cite{nr6}, of fcc Ca and Sr~\cite{nr6,nr7},
TiSi-family ~\cite{nr8}, 3D carbon allotropes ~\cite{nr1},
antiperovskite Cu$_3$(Pb,Zn)N ~\cite{nr2,nr3}, Ca$_3$P$_2$
~\cite{nr4}, photonic crystals ~\cite{Lu.L2013}, and a
hyperhoneycomb lattice ~\cite{nr5}, (\emph{ii}) nodal chain or nodal
net (Fig.~\ref{fig:nl}(b)) in which several closed nodal rings touch
at a point to form a chain (or a net) stretching in the whole BZ of
both iridium tetrafluoride ~\cite{nc1}, HfC ~\cite{nc2}, graphene
network structure~\cite{nc3} and TiS\cite{Chen2017}, (\emph{iii})
nodal link (Fig.~\ref{fig:nl}(c)) featured with closed nodal loops
which are linked and interwined with each other in so-called Hopf
link semimetal Co$_2$MnGa ~\cite{nl1} and (\emph{iv}) nodal knots
(Fig.~\ref{fig:nl}(d)) in which a nodal ring entangles with itself
to form nodal knot structure~\cite{nk1}. Besides these four main
types of nodal lines structures, some discontinuous nodal lines can
also exists in BZ, such as in monoclinic C$_{16}$ ~\cite{nls1} and
metallic rutile oxides ~\cite{nls2}. To date, no Weyl nodal
\emph{straight} lines (Fig. ~\ref{fig:nl}(f)) have been reported in
various materials.

Within this context, through first-principles calculations, we
report that the famous high-temperature BCS-type
superconductor\cite{HSC1,HSC2}, MgB$_2$, hosts the phononic
topological Weyl nodal lines (PTWNLs) in its bulk BZ of the phonon
spectrum. MgB$_2$ has a AlB$_2$-type hexagonal lattice \cite{Str1}
in a space group of $P6/mmm$ (No. 191) in Fig.
\ref{fig:structure}(a). Mg at the Wickoff site $1a$ (0,0,0) and B
occupying the $2d$ ($\frac{1}{3}$, $\frac{2}{3}$, $\frac{1}{2}$)
site form two-dimensional triangular and honeycomb lattices,
respectively. Two layers interleaved along $z$-axis. Based on
symmetry analysis, it has been noted that there are three
double-degenerate phonon bands along the directions from
$-H(\frac{1}{3},\frac{1}{3},-\frac{1}{2})$ to
$H(\frac{1}{3},\frac{1}{3},\frac{1}{2})$ in the bulk BZ as
illustrated in Fig. \ref{fig:structure}(b). The calculations
demonstrate that these double-degenerate phonon bands exactly are
PTWNLs. The Berry phases of the defined loops around these PTWNLs
are derived to be $\pm\pi$, indicating their non-trivial topological
nature. Berry curvature distributions uncover their chirality. In
addition, on the $(10\bar{1}0)$ surface, we observe PTWNLs-induced
drumhead-like non-trivial surface states, which are indeed very
similar with the topological surface states found in topological
Dirac nodal line systems. When the inversion symmetry is broken, the
double-degenerate PTWNLs can be further lifted up and WPs can be
formed. Afterwards, the surface states form opening Fermi arc
structure connecting these WPs. In addition, we have noted that the
other three doubly-degenerate phononic bands form $-$$A(0, 0,
-\frac{1}{2})$ to $A(0, 0, \frac{1}{2})$ in the bulk BZ are
topologically trivial, without appearance of any phononic
topological surface states.

\emph{Methods.---} We perform first-principles calculations based on
density functional theory (DFT) \cite{DFT1,DFT2} with generalized
gradient approximation (GGA) in the form of Perdew-Burke- Ernzerhof
function \cite{PAW1,PBE} for exchange-correlation potential. The
calculations are performed using the Vienna \emph{ab initio}
simulation package (VASP) \cite{VASP1,VASP2}. A self-consistent
field method (tolerance $10^{-5}$ eV/atom) is employed in
conjunction with planewave basis sets of cutoff energy of 500 eV.
Atomic structure optimization is implemented until the remanent
Hellmann- Feynman forces on the ions are less than 0.001 eV/{\AA}.
We use a $\Gamma$-centered $11\times11\times17$ k point mesh to
sample the BZ. Phonon spectra are obtained using the density
functional perturbation theory (DFPT) with a Born-Karman boundary
condition implemented in the Phonopy package~\cite{Phonopy}. The
force constants are calculated using a $3\times3\times3$ supercell
and are taken as the tight-binding parameters to build the dynamic
matrices. The surface DOS are obtained by using the iteration
Green's function method~\cite{Green}.

\emph{Strucuture.---} The optimized lattice constants are $a =
3.074~{\AA}$ and $c = 3.513~{\AA}$. These values are in good
agreements with previous experimental~\cite{HSC1,HSC2,Str1} and
computational~\cite{Strc1,Strc2,Strc3,Strc4,Strc5} results. The
structure is shown in Fig.~\ref{fig:structure}(a).

\emph{Phonon spectra.---} Figure~\ref{fig:structure}(c) presents the
phonon spectrum of MgB$_2$ along high symmetry momentum paths, which
are in nice agreement with previous calculations
\cite{Strc1,Strc2,Strc3,Strc4,Strc5} and experimental
characterizations\cite{Exp1,Exp2,Exp3}. At the high-symmetry
$K(\frac{1}{3},\frac{1}{3}, 0)$ and
$H(\frac{1}{3},\frac{1}{3},\frac{1}{2})$ points, there are three
double-degenerate points each.
%This is due to the little group at
%both $K$ and $H$ has three 1-dimensional and also three
%2-dimensional representations.
Phonon spectrum along $-H$ to $H$ in
Fig.~\ref{fig:structure}(c) evidences the existence of the three
double-degeneracy bands, which are PTWNLs in the strictly
\emph{straight} lines traversing the whole BZ. In
Figs.~\ref{fig:structure}(d) and~\ref{fig:structure}(e), we depicted
the phonon dispersions along a PTWNL on the $xz$-plane and around a
specified point, (1/3,1/3,0.2) in an unit of reciprocal lattice
vector, of a PTWNL on the $xy$-plane. It can be seen that their
dispersion around any point of the PTWNLs exhibits a linear
dispersion on the $xy$-plane. It is a hallmark of a WP. In other
words, these 1st, 2nd, and 3rd PTWNLs  in Fig.
\ref{fig:structure}(c) are highly unique. In the first, they along
the $-$H-K-H direction can be viewed as countless WPs closely
aligned in the straight lines in its three-dimensional BZ. In the
second, each PTWNL can extend through the whole BZ along this
high-symmetry direction in the momentum space. In particular, all of
them exactly lie at the BZ boundary along the one-way high-symmetry
direction, which are indeed protected by the mirror symmetry. In
addition, it needs to be emphasized that, although such Dirac-line
behaviors stretching the whole BZ along the one direction has been
recently outlined in the electronic structures of rhombohedrally
stacked honeycomb lattices \cite{Hyart} and of nanostructured carbon
allotropes \cite{Zhangnanolett}, these symmetry-protected PTWNLs
along the one-direction high-symmetry lines in the phonon spectrum
of MgB$_2$ is still recognized for the first time. Mechanically,
these three different PTWNLs are associated with three distinct
phonon modes. The 1st PTWNL mainly involves the in-plane vibrational
mode of Mg atoms, the 2nd one is totally determined by the
out-of-plane vibrational mode of two boron atoms moving in opposite
or same directions along the $c$ axis while the Mg is stationary,
and the 3rd one only involves the in-plane motion in which boron
moves along the \emph{x} or \emph{y} axes and Mg along almost the
diagonal direction of the \emph{xy} axes. Another difference for
these three PTWNLs is that the 2nd PTWNL shows an apparent
$k$-dependent dispersion of the frequencies, whereas both the 1st
and 3rd PTWNLs are nearly flat in the frequencies against \emph{k}
vector along the $-$H-K-H or $-$H$^\prime$-K$^\prime$-H$^\prime$
directions as shown in Fig. \ref{fig:structure}(c).

\emph{Topological properties of phonon.---} To determine the
topological nature of these PTWNLs, we have derived the
corresponding Berry phases and Berry curvature distributions. For a
closed loop in 3D BZ, the berry phase is defined as
\begin{equation}
\gamma_n = \oint_C \mathbf{A}_n(\mathbf{k}) \cdot d\mathbf{l},
\end{equation}
where $\mathbf{A}_n(\mathbf{k}) = i\langle u_n(\mathbf{k}) |
\nabla_k |u_n(\mathbf{k}) \rangle$ is the Berry connection and
$u_n(\mathbf{k})$ is the Bloch wavefunction of $n$-th band. As
illustrated in Fig. \ref{fig:structure}(b), we have first selected a
closed circle on the $xy$-plane centered at a momentum position of
the PTWNL (see the circle marked by the green curves). Note that the
radius of the closed circles going around each PTWNL can be selected
to be arbitrary large, as long as it does not also cover another
PTWNL. Interestingly, we have found that their Berry phase for all
three PTWNLs along the $-$H-K-H direction are $\pi$, whereas the
other three PTWNLs along the $-$H$^\prime$-K$^\prime$-H$^\prime$
direction have an opposite Berry phase of $-$$\pi$. This means that
these PTWNLs are topological non-trivial and it also proves that the
topological nature of the PTWNLs along the $-$H-K-H direction and
$-$H$^\prime$-K$^\prime$-H$^\prime$ direction are opposite in their
chirality. As a further supporting evidence, we can directly
visualize the source or sink effects by deriving the Berry curvature
distributions, $\mathbf{\Omega}_n(\mathbf{k}) = \nabla \times
\mathbf{A}_n(\mathbf{k})$, on the $xy$-plane, as shown in
Figure~\ref{fig:structure}(f).

In addition, as shown in the phonon spectrum from $\Gamma$ to A in
Fig. \ref{fig:structure}(c) there are also three two-fold degenerate
lines. However, the analysis of their Berry phases and Berry
curvatures confirm that these three lines are indeed topological
trivial. Their associated vibrational phonon modes were already
discussed in a previous publication~\cite{Strc5}.

\begin{figure}[hbt]
\centering \vspace{0.1cm}
\includegraphics[width=0.45\textwidth]{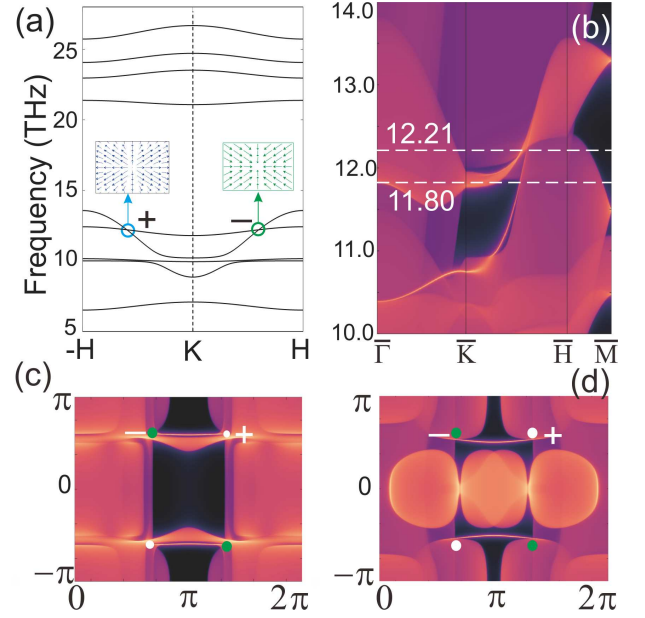}
\caption{Phonon dispersion along the $-$H-K-H direction of the
artificial MgBC and its (10$\bar{1}$0) surface phonon dispersion.
(a) The phonon spectra along high symmetry $-$H-K-H line. Two WPs
are observable with opposite chirality, $+$ and $-$, as illustrated
by Berry curvature distributions in the insets. (b) The surface
phonon spectrum along the high-symmetry lines in the (10$\bar{1}$0)
surface BZ to show the non-trivial surface states. Two dashed lines
denote the frequencies of 11.80 THz and 12.21 THz, respectively.
Opening arc states connecting a pair WPs with opposite chirality in
(c) the surface Fermi-like plot with the frequency of 12.21 THz and
(d) the surface Fermi-like plot with the frequency of 11.80 THz.  }
\label{fig:smb}
\end{figure}

\emph{Phononic drumhead-like surface states.---} Furthermore, we
have identified the surface phonon spectrum of the (10$\bar{1}$0)
surface of MgB$_2$. As expected, there should have topologically
protected phononic non-trivial drumhead-like surface states in the
rectangular region confined by the projected lines of the PTWNLs
along the K to H direction and along the $K^\prime$ to $H^\prime$
direction. We calculate surface phononic density of states (PDOSs)
by using the iteration Green's function method~\cite{Green}. After
Fourier transforming the force constants, surface Green's function
is built from dynamical matrix and the surface DOS is taken from the
imaginary part of surface Green's function. Figure~\ref{fig:ss}
presents the surface states on the MgB$_2$ $(10\bar{1}0)$ surface.
In Fig.~\ref{fig:ss}(a), we depict the surface PDOSs along high
symmetry moment paths on the surface BZ ranging from 12.0 THz to
16.0 THz to clearly show the phononic surface states associated with
the 2nd PTWNL. It can be seen in Fig. \ref{fig:ss} that in the
$\bar{X}$-$\bar{M}$ direction the bright phononic non-trivial
surface states are clearly observable. In Figs.~\ref{fig:ss}(b) and
~\ref{fig:ss}(c), we further show the evolutions of drumhead-like
surface states with respect to increasing $k_z$ and frequency,
respectively. We observe that the surface states are confined well
in the rectangular region outlined by the projected lines of two 2nd
PTWNLs with opposite chirality. In Fig. \ref{fig:ss}(b), with
increasing $k_z$ the downward parabolic-shape non-trivial surface
states climb up to a higher frequency and the depth of the parabola
becomes smaller and smaller. At $k_z = 0.5$ in Fig. \ref{fig:ss}(b),
the shallow parabolic non-trivial surface state is completely above
the surface state at $k_z = 0.0$. This leads to an interesting
evolution of the phononic non-trivial surface states as the
frequency increases. We have observed three typical types of
non-trivial surface states associated with the 2nd PTWNL in
Fig.~\ref{fig:ss}(c). When the frequency $13.3 \lesssim f \lesssim
13.9$ THz, the surface states looks like to connect two points on
the same projected line of the bulk PTWNL. When $ 13.9 \lesssim f
\lesssim 14.8$ THz, the surface states clearly connect two points on
the two projected lines of the bulk PTWNLs with opposite chirality.
When $ 14.8 \lesssim f \lesssim 15.1$ THz, the surface states form a
ring on the 2D surface BZ and do not intersect with any projected
lines of the bulk PTWNLs. Indeed, in another two different frequency
regions below 12 THz or above 16 THz, we can certainly observe the
phononic non-trivial surface states which are associated with the
1st and 3rd PTWNLs along the bulk K-H directions.

\emph{Phononic Weyl points and surface arcs.---} Interestingly, our
calculations also reveal that the double detergency of these PTWNLs
can be lifted, possibly resulting in the appearance of WPs, when
both the mirror and inversion symmetries are broken. To elucidate
this feature, we have substituted one boron atom with a carbon atom
in the MgB$_2$ unit cell, leading to the composition of MgBC. With
such a treatment, the MgBC structure now belongs to the space group
of $P\bar{6}m3$ (No. 187) because the inversion symmetry is broken
and, simultaneously, the mirror symmetry of the K-H direction is
also broken. As illustrated in Fig. \ref{fig:smb}(a), the derived
phonon dispersion along the H to K direction does not exhibit any
PTWNLs due to the lifting up of double degeneracy. In contrast, a
pair WPs appear because of the phonon band crossings. We also
analyzed their topological nature using the Berry curvature
distributions around each WP, obtaining the nonzero topological
charges of $\pm$1 as shown in Fig. \ref{fig:smb}(a). Additionally,
on the $(10\bar{1}0)$ surface we have clearly observed the opening
phononic surface arc states connecting to such a pair WPs with
opposite chirality in Fig. \ref{fig:smb}(b,c and d).

Finally, we have found that a series of metallic diborides, which
are isostructural to MgB$_2$ in the AlB$_2$ family, host similar
PTWNLs in their phonon spectra. These materials include the
prototype material AlB$_2$, BeB$_2$, CaB$_2$, CuB$_2$, NaB$_2$,
OsB$_2$, ScB$_2$, SrB$_2$, TaB$_2$, TiB$_2$, YB$_2$ and ZrB$_2$. For
each of them, there are three double-degenerate PTWNLs along $-$H to
H in the 3D BZ. These PTWNLs transverse the whole BZ and act as
source or sink of the Berry curvatures. On their (10$\bar{1}$0)
crystal surface, we observed drumhead-like non-trivial phononic
surface states. At certain frequencies, the drumhead-like surface
states form isolated arc structures and should serve as clear
signature for experimental verifications. Available experimental
techniques \cite{neu-s,eels-1,eels-2,Xray} to probe these states
include neutron scattering, electron energy loss spectroscopy and
X-ray thermal diffuse scattering.

\emph{Note added.---} During the preparation of this work, we became
aware of a very recent manuscript proposing the electronic
topological Dirac nodal lines in MgB$_2$ through first-principles
calculations~\cite{DNLs}.

\emph{Acknowledgements} Work was supported by the National Science
Fund for Distinguished Young Scholars (No. 51725103), by the
National Natural Science Foundation of China (Grant Nos. 51671193
and 51474202), and by the Science Challenging Project No. TZ2016004.
All calculations have been performed on the high-performance
computational cluster in the Shenyang National University Science
and Technology Park and the National Supercomputing Center in
Guangzhou (TH-2 system) with special program for applied research of
the NSFC-Guangdong Joint Fund (the second phase) under Grant
No.U1501501.

\bigskip
\noindent $^{\textcolor{blue}{\ddagger}}$ These authors contributed
equally to this work.

\end{document}